\documentstyle[12pt,aps]{revtex}
\begin{document}
\vskip 5cm
\title{\ \\\ \\\ \\\
On Solar Model Solutions to the Solar Neutrino Problem}
\author{X. Shi$^a$, D. N. Schramm$^{a,b}$ and D. S. P. Dearborn$^c$\\
\vskip 0.1cm
$^a$The University of Chicago, Chicago, IL 60637-1433\\
$^b$NASA/Fermilab Astrophysics Center, Fermi National Accelerator Laboratory,\\
Box 500, Batavia, IL 60510-500\\
$^c$Lawrence Livermore National Laboratory, Livermore, CA 94550}
\maketitle
\vskip 0.3cm
\begin{abstract}
{Without assuming any solar models and neutrino flavor conversions, $^8$B
neutrinos seen by the Kamiokande
experiment should contribute 2.6$\pm 0.45$ SNU to the chlorine
experiment. When this rate is compared with the
total event rate of 2.3$\pm 0.2$ SNU observed
by the Homestake experiment which should inlude a 0.2 SNU contribution
from uncertainty-free pep neutrinos,
there may still be a possible evidence that $^7$Be neutrinos are more
severely suppressed than the $^8$B neutrinos with respect to the predictions
of standard solar models, which cannot be explained by any known
astrophysics solution.
Given a Kamiokande event rate of larger than
36$\%$ (2$\sigma$) of the prediciton of Bahcall and
Pinsonneault's standard solar model, variations of standard solar models
yield minimal rates of 3.6 SNU for the Homestake experiment and
114 SNU for GALLEX and SAGE. Therefore, variations of standard
solar models as solutions to the solar neutrino problem are so far
inconsistent with the Homestake experiment and only marginally allowed by the
gallium experiments.
If the gallium experiments confirm a flux significantly below 114 SNU,
it would imply unconventional physics.}

\end{abstract}
\noindent{-----------------------------------------------\\}
\noindent{Submitted to {\sl Phys. Rev.} D}
\vfill\eject
\section{Introduction}
The observed deficit of the solar neutrino flux with
respect to the prediction of the standard solar models
has been one of the most interesting problems in both astrophysics and particle
physics \cite{BP,Turck,Langacker,ShiSchramm}.
The so-called ``Solar Neutrino Problem'' received renewed
interest recently following reports of new data possibly conflicting
with standard assumptions on the crucial $^7$Be(p,$\gamma$)$^8$B reaction rate
\cite{Motobayashi,Caltech} which led to renewed attempts to
solve the problem by modifying the standard solar
models \cite{ShiSchramm,DarShaviv}.
In this letter, we attempt to review the current
solar neutrino situation in the light of the most recent nuclear experimental
results and solar neutrino experimental results, and to examine
the minimal rates that successful solar models can yield for
the Homestake and two gallium experiments.
In particular, we note that even with the assumption of a very low
 $^7$Be(p,$\gamma$)$^8$B rate, the Homestake experiment is still in
conflict with variations of standard solar models.
We also showed that the gallium experiments
will also be in conflict if they are proven to have fluxes
significantly below 114 SNU (1 SNU=$10^{-36}$ capture/target atom/second).

The observed solar neutrino flux of four currently available
solar neutrino experiments, the Homestake $^{37}$Cl capture
experiment \cite{Homestake}, the Kamiokande $\nu$-$e$
scattering experiment \cite{KAM},
the GALLEX $^{71}$Ga capture experiment \cite{GALLEX}
and the SAGE $^{71}$Ga capture experiment \cite{SAGE}, have been summarized in
Table 1. All of them are lower than the predictions of
the most referenced standard solar models of
Bahcall and Pinsoneault (BPSSM) \cite{BP},
or the standard solar model of Turck-Chi\`eze {\sl et al.}
(TLSSM) \cite{Turck}.
It should be noted that alternative solar model calculations give essentially
identical results if the same input parameters (nuclear reaction rates,
radiative opacities, the heavy element abundance in the sun,
the age of the sun, etc.) are used \cite{BP}.

Among the four experiments, only the Kamiokande experiment has been fully
calibrated. The two gallium experiments, GALLEX and SAGE,
although their initial results were seemingly inconsistent
\cite{SAGE1,GALLEX1},
are increasingly consistent with each other after more than two years of
observations.
The Homestake experiment, which observed the most significant solar neutrino
deficit with respect to the standard solar models and is therefore crucial
to the solar neutrino problem, naturally draws questions since so much is
currently relying on the experiment. Although no serious questions about
the experiment have been proven, some still worry that
its event rate seems to be inconsistent with
a constant solar neutrino flux, in particular its event rate
after 1985--1986 pump failures (2.8$\pm 0.3$ SNU \cite{Lande})
is significantly higher than that before (2.1$\pm 0.3$ SNU \cite{Davis}).
The latter point was elaborated by some attempts to reconcile the solar
neutrino experiments and
the standard solar models, arguing that a higher Homestake rate after
1986 might not be far below some modified standard solar
models \cite{DarShaviv,Morrison}.

However, it should be pointed out that the deviation of the Homestake
data from a constant rate is only marginally significant \cite{Bahc,Fili,SSRD}
and there is no ground besides different rates that discriminates
data before 1984 from those after 1986. Different pump configurations
and different pumping times have
been introduced yielding no obvious change in
the capture rate \cite{NA}. Various tests by the Davis group
also showed no unexpected systematic uncertainties \cite{NA}.
Furthermore, the rate
after 1986 is still statistically consistent with the rate before 1984 within
about 2$\sigma$. It is therefore completely unjustified to use only
the Homestake data after 1986 in discussing the solar neutrino
problem since if the rate did
shift, then there are unexpected systematic uncertainties in the experiment,
which put the whole experiment into question until it can be resolved.

Attempts to lower the solar neutrino flux predicted by solar models
appropriately concentrate on two approaches: (1) lowering
the core temperature $T_{\rm c}$ of the Sun;
(2) using a lower $^7$Be(p,$\gamma$)$^8$B rate, because the uncertainties
from these two factors are known to be
far greater than the uncertainties from the other
aspects of solar models \cite{NA}. In the context of the standard
solar models (i.e., no rotations, no magnetic fields, no exotic particles,
and standard nuclear reaction network, etc. \cite{NA}),
a lower $T_{\rm c}$ can be achieved by a lower heavy element abundance of the
sun, $Z$, and/or lower radiative opacities at the center of the sun.
Most if not all non-standard solar models, such as
invoking assumptions like a
10$^9$ Gauss magnetic field in the solar interior, or a black hole
at the center of the sun, or captures of Weakly
Interacted Massive Particles (WIMPs) by the sun, or even invoking
additional core mixing, also end up reducing the neutrino flux by
effectively lowering the core temperature $T_c$ of the sun \cite{NA}.
They can be therefore categorily included in the first approach to
lower the predicted solar neutrino flux.

It has been shown by Bahcall {\sl et al.}
that in standard solar models, the predicted
fluxes from different neutrino sources
depend very differently on the core temperature of the models \cite{BU,NA}.
For example, for pp neutrinos, $^7$Be neutrinos and $^8$B neutrinos,
which are predicted to constitute most of the neutrinos detected
in solar neutrino experiments,
\begin{equation}
\phi ({\rm pp})\propto T_{\rm c}^{-1.2},\quad
\phi (^7{\rm Be})\propto T_{\rm c}^8,\quad
\phi (^8{\rm B})\propto T_{\rm c}^{18},
\end{equation}
where $\phi$'s are neutrino fluxes.
Therefore, a $\pm 1.5\%$ variation in $T_{\rm c}$ alone, which is readily
achievable by adjusting standard solar models within a reasonable range
(for example, the $\pm$10$\%$ 1$\sigma$ uncertainty in $Z$ alone
changes $T_{\rm c}$ by $\pm$0.8$\%$, as our calculations
show approximately that $Z\propto T_{\rm c}^{0.08}$), will
result in a variation in the $^8$B solar neutrino flux by a factor of 2. On the
other hand, the resultant $^7$Be solar neutrino flux only varies by
less than 30$\%$,
while the predicted pp neutrino flux is essentially free of uncertainties
from a variation in $T_{\rm c}$ in standard solar models.

The uncertainty in the $^7$Be(p,$\gamma$)$^8$B rate brings
significantly more uncertainty
to the $^8$B neutrino flux. This rate linearly affects the $^8$B neutrino
flux (since the $^8$B neutrinos are the decay product of the resultant $^8$B
excited state). Its uncertainty comes from both theoretical extrapolations and
interpretations of experimental data themselves \cite{Johnson}. Theoretical
extrapolations for six previous experiments yielded $S_{17}(0)$ for the
reaction ranging from 16 to 42 eV$\cdot$barn, with an weighted average
of $22.4\pm 2.1$ eV$\cdot$barn \cite{Johnson}.
($S(E)=\sigma (E) E{\rm exp}(-2\pi\eta)$, where $\sigma (E)$ is
the cross section. $\eta=e^2Z_1Z_2/\hbar v$ where
$Z_1$ and $Z_2$ are the charges
of colliding particles and $v$ is their relative velocity. The subscript
17 denotes reaction $^7$Be(p,$\gamma$)$^8$B.)
Only two of the six experiments, namely those of
Kavanagh (1969) \cite{Kava69} and Filippone {\sl et al.}
(1983) \cite{Fili83}, went
to energies well below the $M1$ resonance at $\sim$0.6 MeV. They yield
25.2$\pm 2.4$ eV$\cdot$barn (Kavanagh) and 20.2$\pm 2.3$ eV$\cdot$barn
(Filippone) respectively, which disagree with each other at 2$\sigma$.
Several favor the experiment of
Filippone {\sl et al.} since it is the latest among the two and
was published in a refereed journal \cite{ShiSchramm,Morrison}.
The preliminary result of a new experiment by Motobayashi {\sl et al.},
which avoids the $M1$ resonance by using the
reverse reaction of disassociating $^8$B by the
coulomb field of heavy nuclei, implies a low $S_{17}$(0).
A new extrapolation by Langanke and Shoppa, which argues for a large $E2$
contribution to the rate,
if proven correct, could imply $S_{17}(0)$ as low as 12$\pm 3$eV$\cdot$barn.
Such a low $S_{17}(0)$ would obviously be in conflict with previous
measurements. However, given the current situation,
one should keep an open mind and possibly
allow $S_{17}$ to vary downward by as much as a factor of 2 from the previous
average.

\section{Implications of Solar Neutrino Experiments}
Given such uncertainties in the $^8$B neutrino flux,
none of the individual solar neutrino experiments taken by themselves
indicates strong evidence for a solar neutrino deficit that
cannot be resolved with solar model variations:
the Kamiokande experiment observes
$^8$B neutrinos only, hence its event rate may serve better
as a normalization of the absolute $^8$B neutrino flux
than an indicator of a solar neutrino deficit;
the Homestake experiment, which should observe mostly $^8$B neutrinos plus some
$^7$Be neutrinos (and a few pep neutrinos and CNO neutrinos),
is also subject to a large uncertainty in its expected flux;
the two gallium experiments, although capable of observing
the uncertainty-free pp neutrinos, see a solar neutrino flux
that is larger than the pp neutrino flux, hence fail to show a deficit
in the model independent pp neutrinos.

Nevertheless, a problem for solar model solutions to the solar neutrino
problem may still persist if results from the Kamiokande experiment and the
Homestake experiment are combined \cite{ShiSchramm,BB,Bludman,Berezinski}.
That is, if one normalizes the $^8$B neutrino flux predicted by
solar models to the Kamiokande result,
the $^8$B neutrinos should still contribute 3.1$\pm 0.4$ SNU
to the Homestake result.
(An argument that the two experiments with different energy
thresholds may observe different reductions in the $^8$B neutrinos with respect
to a standard solar model immediately implies a departure in the
$^8$B neutrino spectrum from that predicted by the
standard electroweak theory, and thus new particle physics.)
The observed Homestake rate of 2.3$\pm 0.2$ SNU, therefore,
indicates that the $^7$Be neutrinos which should also be seen by the Homestake
experiment, suffer more reduction than the $^8$B neutrinos with respect to
a particular standard solar model, in this case, BPSSM.

A similar but model-independent argument is also intriguing.
With a lower neutrino energy threshold
the Homestake experiment should see all the neutrinos observed by
the Kamiokande (namely $^8$B neutrinos)
and some neutrinos that have energies below the Kamiokande
threshold (namely $^7$Be neutrinos, pep neutrinos and CNO neutrinos).
Without any solar model assumption on the
solar neutrino spectrum and flux, and with the assumption of
no neutrino oscillation, solar neutrinos observed by the Kamiokande
should contribute 2.6$\pm 0.45$ SNU to the Homestake result.
Figure 1(a) shows a neutrino spectrum that yields the best fit to the
Kamiokande data and 2.6 SNU to the Homestake experiment, and
a neutrino spectrum that is excluded by the Kamiokande result at
$95\%$C.L. and yields 1.7 SNU to the Homestake experiment. Figure 1(b)
shows their resultant recoil-electron spectra compared with
the spectrum observed
by the Kamiokande experiment. All spectra in figure 1
are normalized by the prediction of BPSSM.
The pep neutrinos whose flux is directly tied to the pp neutrino flux, which is
essentially uncertainty-free, are insensitive to solar model
uncertainties and should contribute another 0.2 SNU to the Homestake
result \cite{NA}.
Therefore, the Homestake result of 2.3$\pm 0.2$ SNU provides a possible
evidence that the $^7$Be neutrinos are more severely suppressed with respect to
predictions of standard solar models than the $^8$B neutrinos.
As a result, to reconcile the Homestake result and the Kamiokande result
requires either an explantion of a larger reduction in the $^7$Be neutrinos
than that in the $^8$B neutrinos with respect to predictions of standard solar
models, or an assumption of contributions from other neutrino flavors to
the Kamiokande experiment, both of which would imply new neutrino physics.

So far, no modification in solar models with known physics
can explain a more severe reduction
in $^7$Be neutrinos than in $^8$B neutrinos with respect to
BPSSM \cite{ShiSchramm,BB,Bludman,Berezinski}.
Lowering $T_{\rm c}$ in standard solar models
will only suppress more $^8$B neutrinos since they are more
$T_{\rm c}$-sensitive, as seen from eq. (1). This conclusion is also valid
for non-standard solar models, since the $^8$B neutrinos are intrinsically
more $T_{\rm c}$ dependent than the $^7$Be neutrinos due to a higher
coulumb barrier in the reaction that produces the $^8$B neutrinos \cite{NA}.
In terms of nuclear reactions, the $^7$Be(e,$\nu_{\rm e}$)$^7$Li reaction
that produces $^7$Be neutrinos and the $^7$Be(p,$\gamma$)$^8$B reaction
that produces $^8$B neutrinos are the two branches of the $^7$Be reactions
in the sun, with branching ratios of 99.6$\%$ and 0.4$\%$
respectively\cite{NA}.
The only artificial way to achieve a greater reduction in the $^7$Be neutrinos
than in the $^8$B neutrinos is then to suppress the $^7$Be production
rate (by either suppressing $^3$He($^4$He,$\gamma$)$^7$Be rate at low energy
by a least a factor of 2 or increasing the $^3$He($^3$He,2p)$^4$He
rate at low energy by more than a factor of 4 to
account for the $^8$B neutrinos observed by the Kamiokande
\cite{Castellani}) which affects
the $^7$Be neutrinos and $^8$B neutrino equally,
and at the mean time increase the $S_{17}(0)$ or raise $T_{\rm c}$.
But besides the fact that $S_{34}(0)$ and $S_{33}(0)$
(which are astrophysical factors similarly defined as $S_{17}(0)$ but for the
$^3$He($^4$He,$\gamma$)$^7$Be reaction and the
$^3$He($^3$He,2p)$^4$He reaction) are
currently determined to within about 10$\%$ and 20$\%$ respectively \cite{BP},
$S_{17}(0)$ is currently shown
both experimentally and theoretically
to have a downward uncertainty instead of an upward uncertainty, and
a higher $T_{\rm c}$ will increase CNO neutrinos (which have an even higher
$T_{\rm c}$ dependency than the $^8$B neutrinos) significantly
to escalate the conflict between the Homestake result and the Kamiokande
result rather than solve it.

\section{Variations of Standard Solar Models}
To illustrate the above arguments and to
show to what extent the solar neutrino prediction of
standard solar models can vary,
we construct a series of standard solar models with different
$T_{\rm c}$'s and allow $^7$Be(p,$\gamma$)$^8$B to vary freely.
These models are constructed with Dearborn's solar code \cite{Dearborn} using
the Livermore OPAL opacity table and nuclear reaction rates from
Caughlan and Fowler (1988) \cite{Fowler} except for a freely varying
$^7$Be(p,$\gamma$)$^8$B rate.
As a benchmark, we compare one of our models with Bahcall and Pinsonneault's
no diffusion model (BPSSM w$/$o diffusion) in Table 2. They yield quite
similar results. To see the uncertainties caused by the
$^3$He($^4$He,$\gamma$)$^7$Be rate and the
$^3$He($^3$He,2p)$^4$He rate,
we also calculated solar models with different $S_{34}(0)$ and $S_{33}(0)$,
although the variations in
solar neutrino fluxes from their uncertainties are relatively small.

Models with different $T_{\rm c}$ are
constructed by varying $Z$ between 0.014 and 0.021. The measured
value of $Z$ is 0.0245 times the solar hydrogen abundance, or roughly
$0.0177\pm 0.0017$ \cite{Grev}. Our
lowest $T_{\rm c}$ model has a $Z$ of 0.014, that is more than $2\sigma$
below the measured value. Models with such a low $Z$
show distinctive structure differences from models with
$Z\sim 0.018$. For example, the model with $Z=0.014$ has a convective
zone with a depth between 0.720$R_\odot$ and 0.730$R_\odot$
(where $R_\odot$ is the radius of the sun),
much shallower than the measured 0.713$\pm 0.003 R_\odot$
from helioseismology \cite{Christensen}, whereas models
with $Z$ in the range of 0.015 to 0.021 have convective zones
as deep as between 0.718$R_\odot$ and 0.705$R_\odot$.
Models with $Z>0.021$
may also be compatible with helioseismic results but they yield higher
neutrino fluxes contrary to the direction of solving the solar neutrino
problem.

Figure 2 shows predictions of these solar models with
different $T_{\rm c}$ and $S_{17}(0)$ for the four solar neutrino
experiments. Clearly, no overlap region exists between any two experiments
at the $2\sigma$ level for each experiment, except for
a small overlap between the Homestake and SAGE at low $T_{\rm c}$ and low
$S_{17}(0)$. The gap between the Homestake experiment and
the Kamiokande experiment is significantly large, not surprisingly as argued
before, due to the additional contributions from $^7$Be neutrinos,
CNO neutrinos and pep neutrinos in the Homestake experiment. In fact,
as long as the Homestake experiment result is significantly below 3.6 SNU,
a conflict between the Homestake experiment and the
Kamiokande experiment cannot
be solved by simply lowering $T_{\rm c}$ and the
$^7$Be(p,$\gamma$)$^8$B rate in the standard solar models.

The gaps between gallium experiments and the other two experiments
are not yet as severe as the gap between the Homestake experiment
and the Kamiokande experiment. But they are still problematic
for the variations to the standard solar models as we discussed here,
if the gallium experiment rate is significantly less than 114 SNU.
Current gallium experiment results, therefore, may still be marginally
compatible with variations of standard solar models. It will be very
interesting to see how GALLEX rate looks after the calibration of GALLEX
with $^{51}$Cr this year. As statistics improves it may be possible that
gallium experiments like the Homestake chlorine experiment will not be
compatible with a standard solar model normalized by the Kamiokande
result and hence would suggest unconventional physics.

Figure 3 shows predictions of standard solar models with different $T_{\rm c}$
and artificially varied $S_{34}(0)$ and $S_{33}(0)$. $S_{17}(0)$ is set
to be 20 eV$\cdot$barn. Obviously 114 SNU is also
the minimal gallium capture rate that can be
reached by the standard solar models when allowing
$S_{34}$ and $S_{33}$ to vary within their uncertainty
ranges and allowing $T_{\rm c}$ to vary within the constraint from
helioseismology. The gap between the Homestake experiment
and the Kamiokande experiment cannot be narrowed even with a wild
variation of $S_{34}(0)$ by a factor of 2 or $S_{33}(0)$ by
a factor of 4.

Besides lowering $Z$ as we did in our solar models,
lower $T_{\rm c}$ may also be equivalently achieved
by lowering the overall opacities,
increasing the pp reaction rate, or shortening the
age of the sun \cite{Italian}. It is interesting to note
that models with a $Z\ga 0.015$ and
opacities further artificially lowered only at the center (which might
be possible under certain hypotheses \cite{Marx})
may achieve a $T_{\rm c}$ lower
than the lowest $T_{\rm c}$ discussed above and still
satisfy the current helioseismic constraint. But our calculations show
that such models cannot suppress $^7$Be neutrinos and CNO neutrinos
as efficiently as the simple low $Z$ models when a similar $T_{\rm c}$ is
achieved. In addition, the opacities at the solar core can only be
artificially lowered to the extent that the resultant helium abundance
in the sun is $\ga 0.26$ \cite{Anders89}.
As a result, after normalization by the 2$\sigma$ lower limit of the
Kamiokande, the minimal rates predicted by these models
for the Homestake experiment and the
gallium experiments remain roughly the same as we discussed above.

The result we obtained should also hold for models that include
helium diffusion. A comparison between Bahcall and
Pinsonneault's helium diffusion model and no helium diffusion
model \cite{BP} shows that the $T_{\rm c}$ dependency of various
solar neutrino sources (i.e., eq. (1)) remains roughly intact
after consideration of helium diffusion in solar models and the two classes
of models yield similar neutrino fluxes when they have the same
$T_{\rm c}$. Having similar input parameters,
a model with helium diffusion yields a higher $T_{\rm c}$ (and hence higher
$^8$B, $^7$Be and CNO neutrino fluxes) than a model without helium diffusion,
due to a higher helium concentration in the solar core in the diffusion model
that increases the mean molecular weight \cite{BP}.
Therefore, to achieve the same $T_{\rm c}$ of a no diffusion model,
a helium diffusion model has to have a lower $Z$ input
than the no diffusion model. For example,
according to our approximate scaling law of $T_{\rm c}\propto Z^{0.08}$,
a helium diffusion model
that has the same low $T_{\rm c}$ as the $Z=0.014$ no diffusion model
should have $Z\approx 0.0135$. On the other hand,
since the surface helium abundance of a helium diffusion model
is roughly 10$\%$ less than the initial helium abundance
due to gravitational settling \cite{BP}, the constraint
on $Z$ deduced from the measured $Z/X$ for
helium diffusion models becomes stricter than that for no diffusion
models, namely $Z\approx 0.0184\pm 0.0018$ instead of
$Z\approx 0.0177\pm 0.0017$. As a result, a helium
diffusion models with $Z\approx
0.0135$ are more difficult to reconcile with the constraint on $Z$.
It is also questionable if such a low
$Z$ helium diffusion model can satisfy
helioseismic constraints.

As we discussed previously,
all modifications to solar models
that do not contradict known physics
can only suppress the $^7$Be neutrinos by at most the same factor
as they suppress the $^8$B neutrinos with respect to BPSSM. Therefore
in the most extreme cases, if the
Kamiokande experiment sees only 36$\%$ of the BPSSM prediction for $^8$B
neutrinos, the Homestake experiment should then expect 2.7 SNU from the
$^8$B neutrinos and the $^7$Be neutrinos. If we add the 0.2 SNU
contribution from pep neutrinos which has a very small
uncertainty \cite{NA}, and the contribution from
CNO neutrinos, which may be neglected since they
are very sensitive to $T_{\rm c}$ and some nuclear rates,
we expect an absolute minimal rate of about 3.0 SNU for
the Homestake experiment to accomodate any solar model solutions.
Similarly, gallium experiments expect 74 SNU from pp neutrinos
and pep neutrinos, 18 SNU from $^7$Be and $^8$B neutrinos,
and contributions from CNO neutrinos. Therefore, the minimal gallium rate
to accomodate solar models after $^8$B neutrinos being normalized by
the Kamiokande is about 92 SNU if such solar models
can successfully suppress most of the uncertain CNO neutrinos. It should
be noted, however, we have not constructed any realistic solar model
that achieves such extreme reductions.

\section{Summary}
With an overall average Homestake rate
of 2.3$\pm 0.2$ SNU and a Kamiokande rate of 0.50$\pm 0.04\pm 0.06$ times
the BPSSM prediction, there is little space for a convincing
solar model solution to the solar neutrino problem. Variations of standard
solar models yield a minimal rate of 3.6 SNU for the Homestake experiment
and 114 SNU for the gallium experiments, when the $^8$B neutrino flux is
normalized to the Kamiokande result.
For gallium experiments, their current yields haven't been significantly
lower than the minimal 114 SNU. But their accuracy will improve
with time and stronger statements may be possible.

It is nice to know the next generation solar neutrino experiments
(SNO, Super-Kamiokande, Borexino, ICARUS, etc. \cite{NA}) will be able to
distinguish different solutions to the solar neutrino problem within next five
years \cite{SSF}. And the gallium experiments may even be
able to exclude 114 SNU in the not too distant future.
Helioseismic obervations by the on-going
Global Oscillation Network Group (GONG) \cite{GONG} and future Solar and
Heliospheric Observation (SOHO) mission
will also provide much more information on the solar interior,
thus further constrain solar models.

\section{Acknowledgement}
We thank John Bahcall, Sid Bludman, Ray Davis, S. Del'Innocenti,
Moshe Gai, Ken Lande, Naoya Hata, Paul Langacker, Karlheinz Langanke,
Douglas Morrison and Peter Rosen for valuable discussions.
This work is supported by the NASA and the DoE(nuclear) at the University
of Chicago, by DoE at Livermore, and by the DoE and NASA through
grant 2381 at Fermilab.
\vfill\eject

\vfill\eject
\begin{table}
\caption{Solar neutrino fluxes measured by experiments
vs. predictions of standard solar models.\label{table1}}
\begin{tabular}{lcccc}
&Experimental results& BPSSM \cite{BP}& TLSSM \cite{Turck}
&A low flux model \cite{ShiSchramm}\\ \hline
Homestake exp.& 2.3$\pm 0.2$ SNU& 8.0 SNU& 6.4 SNU & 4.7 SNU\\
Kamiokande exp.\tablenotemark[1]& $0.51\pm 0.04\pm 0.06$
 & 1.00 & 0.75 &   0.54\\
Gallium exp.& 79$\pm 10\pm 6$ SNU (GALLEX) & 132 SNU & 123 SNU & 117 SNU\\
&             74$\pm 17\pm 10$ SNU (SAGE)& & & \\
\end{tabular}
\tablenotemark[1]{Normalized by the prediction of BPSSM.}
\end{table}
\vfill
\begin{table}
\caption{Comparison between a model in this work and BPSSM w$/$o
diffusion [1].\label{table2}}
\begin{tabular}{lll}
& BPSSM w$/$o diffusion & A model in this work\\ \hline
Predictions for $^{37}$Cl exp.& & \\
pep neutrinos   & 0.2 SNU & 0.2 SNU\\
$^7$Be neutrinos& 1.1 SNU & 1.1 SNU\\
$^8$B neutrinos & 5.5 SNU & 5.5 SNU\\
$^{13}$N neutrinos& 0.1 SNU & 0.1 SNU\\
$^{15}$O neutrinos& 0.3 SNU & 0.4 SNU\\
Total           & 7.2 SNU & 7.3 SNU\\\hline
Predictions for $^{71}$Ga exp.& & \\
pp neutrinos & 71 SNU & 70 SNU \\
pep neutrinos   & 3 SNU & 3 SNU\\
$^7$Be neutrinos& 34 SNU & 33 SNU\\
$^8$B neutrinos & 12 SNU & 12.4 SNU\\
$^{13}$N neutrinos& 3 SNU & 3.5 SNU\\
$^{15}$O neutrinos& 4 SNU & 6.6 SNU\\
Total & 127 SNU & 129 SNU\\\hline
$Z$ & 0.01895 & 0.0190 \\
$S_{17}(0)$ &22.4 eV$\cdot$b &22.4 eV$\cdot$b\\
$T_{\rm c}$& 1.56$\times 10^7$K & 1.577$\times 10^7$K\\
\end{tabular}
\end{table}
\vfill\eject
\noindent{\bf Figure Captions:}
\medskip

\noindent Figure 1. (a) Neutrino spectrum 1 (the solid line) is
the best fit to the Kamiokande data and yields
2.6 SNU to the Homestake experiment;
neutrino spectrum 2 (the dashed line)
is excluded by the Kamiokande result at
$95\%$C.L. and yields 1.7 SNU to the Homestake experiment. Both
spectra are normalized by the prediction of BPSSM \cite{BP}.
(b) The solid line is the expected spectrum of recoil electrons
from neutrino spectrum 1; the dashed line is
the expected spectrum of recoil electrons
from neutrino spectrum 2. The Kamiokande data are also shown.
All spectra are normalized by the prediction of BPSSM.
\bigskip

Figure 2. It shows the predictions of standard
solar models with different $T_{\rm c}$ and $S_{17}(0)$.
Long-dashed lines: predictions for the Kamiokande experiment,
normalized by the solar model of Bahcall and Pinsonneault \cite{BP};
Solid lines: predictions for the Homestake experiment in unit of SNU;
Short-dashed lines: predictions for gallium experiments in unit of SNU.
Regions allowed by the Kamiokande experiment and
the Homstake experiment at 95$\%$ C. L. are shown by arrows.
The shaded region on the left side is ruled out by the constraint from
helioseismology. The hatched region on the upper center is expected by
standard solar models with common choices of inputs (including their
uncertainties); the rectangular region at the center is expected by
standard solar models with common choices of inputs except
a small $S_{17}(0)$ suggested by ref. 6.
\bigskip

Figure 3. It shows the predictions of standard
solar models with different $T_{\rm c}$, $S_{34}(0)$ and $S_{33}(0)$.
Long-dashed lines: predictions for the Kamiokande experiment,
normalized by the solar model of Bahcall and Pinsonneault \cite{BP};
Solid lines: predictions for the Homestake experiment in unit of SNU;
Short-dashed lines: predictions for gallium experiments in unit of SNU.
The shaded region on the left side is ruled out by the constraint from
helioseismology. The hatched region on the upper center is expected by
standard solar models with common choices of inputs (including their
uncertainties).
\end{document}